\pdfoutput=1
\documentclass[12pt,a4paper]{article}
\usepackage{ifthen} % for conditional statements
\newboolean{pdflatex}
\setboolean{pdflatex}{true} % False for eps figures 

\newboolean{articletitles}
\setboolean{articletitles}{true} % False removes titles in references

\newboolean{uprightparticles}
\setboolean{uprightparticles}{false} %True for upright particle symbols

\newboolean{inbibliography}
\setboolean{inbibliography}{false} %True once you enter the bibliography

\usepackage{microtype}
\usepackage{lineno}  % for line numbering during review
\usepackage{xspace} % To avoid problems with missing or double spaces after
\usepackage{caption} %these three command get the figure and table captions automatically small

\usepackage{graphicx}  % to include figures (can also use other packages)
\usepackage{color}
\usepackage{colortbl}
\graphicspath{{./figs/}} % Make Latex search fig subdir for figures

\usepackage{amsmath} % Adds a large collection of math symbols
\usepackage{amssymb}
\usepackage{amsfonts}
\usepackage{upgreek} % Adds in support for greek letters in roman typeset

\newcommand*\patchAmsMathEnvironmentForLineno[1]{%
\expandafter\let\csname old#1\expandafter\endcsname\csname #1\endcsname
\expandafter\let\csname oldend#1\expandafter\endcsname\csname
end#1\endcsname
 \renewenvironment{#1}%
   {\linenomath\csname old#1\endcsname}%
   {\csname oldend#1\endcsname\endlinenomath}%
}
\newcommand*\patchBothAmsMathEnvironmentsForLineno[1]{%
  \patchAmsMathEnvironmentForLineno{#1}%
  \patchAmsMathEnvironmentForLineno{#1*}%
}
\AtBeginDocument{%
\patchBothAmsMathEnvironmentsForLineno{equation}%
\patchBothAmsMathEnvironmentsForLineno{align}%
\patchBothAmsMathEnvironmentsForLineno{flalign}%
\patchBothAmsMathEnvironmentsForLineno{alignat}%
\patchBothAmsMathEnvironmentsForLineno{gather}%
\patchBothAmsMathEnvironmentsForLineno{multline}%
}

\usepackage{hyperref}    % Hyperlinks in references
\usepackage[all]{hypcap} % Internal hyperlinks to floats.

\def\lhcb {\mbox{LHCb}\xspace}

\ifthenelse{\boolean{uprightparticles}}%
{

 \def\Pmu         {\ensuremath{\upmu}\xspace}

 \def\Ppsi        {\ensuremath{\uppsi}\xspace}

 \def\PDelta      {\ensuremath{\Delta}\xspace}                 
 \def\PXi      {\ensuremath{\Xi}\xspace}                 
 \def\PLambda      {\ensuremath{\Lambda}\xspace}                 
 \def\PSigma      {\ensuremath{\Sigma}\xspace}                 
 \def\POmega      {\ensuremath{\Omega}\xspace}                 
 \def\PUpsilon      {\ensuremath{\Upsilon}\xspace}                 
 
 \def\PB      {\ensuremath{\mathrm{B}}\xspace}                 
                  
 \def\PD      {\ensuremath{\mathrm{D}}\xspace}

 \def\PJ      {\ensuremath{\mathrm{J}}\xspace}                 
 \def\PK      {\ensuremath{\mathrm{K}}\xspace}

 \def\Pi      {\ensuremath{\mathrm{i}}\xspace}

}
{

 \def\Pmu         {\ensuremath{\mu}\xspace}

 \def\Ppsi        {\ensuremath{\psi}\xspace}                 
                  
 \mathchardef\PDelta="7101
 \mathchardef\PXi="7104
 \mathchardef\PLambda="7103
 \mathchardef\PSigma="7106
 \mathchardef\POmega="710A
 \mathchardef\PUpsilon="7107
                  
 \def\PB      {\ensuremath{B}\xspace}                 
                  
 \def\PD      {\ensuremath{D}\xspace}

 \def\PJ      {\ensuremath{J}\xspace}                 
 \def\PK      {\ensuremath{K}\xspace}

 \def\Pi      {\ensuremath{i}\xspace}

}

\def\mup        {{\ensuremath{\Pmu^+}}\xspace}
\def\mun        {{\ensuremath{\Pmu^-}}\xspace} % muon negative (\mum is taken)
\def\mumu       {{\ensuremath{\Pmu^+\Pmu^-}}\xspace}

  \def\Kbar    {{\kern 0.2em\overline{\kern -0.2em \PK}{}}\xspace}

  \def\Dbar    {{\kern 0.2em\overline{\kern -0.2em \PD}{}}\xspace}

\def\Bbar    {{\ensuremath{\kern 0.18em\overline{\kern -0.18em \PB}{}}}\xspace}

\def\jpsi     {{\ensuremath{{\PJ\mskip -3mu/\mskip -2mu\Ppsi\mskip 2mu}}}\xspace}

  \def\Y#1S{\ensuremath{\PUpsilon{(#1S)}}\xspace}% no space before {...}!
\def\OneS  {{\Y1S}}
\def\TwoS  {{\Y2S}}
\def\ThreeS{{\Y3S}}

\def\Lbar        {{\ensuremath{\kern 0.1em\overline{\kern -0.1em\PLambda}}}\xspace}

\def\BF         {{\ensuremath{\cal B}}\xspace}

\def\to                 {\ensuremath{\rightarrow}\xspace}

\def\AT#1     {\ensuremath{A_{\mathrm{T}}^{#1}}\xspace}           % 2

\def\C#1      {\ensuremath{\mathcal{C}_{#1}}\xspace}                       % 9
\def\Cp#1     {\ensuremath{\mathcal{C}_{#1}^{'}}\xspace}                    % 7
\def\Ceff#1   {\ensuremath{\mathcal{C}_{#1}^{\mathrm{(eff)}}}\xspace}        % 9  
\def\Cpeff#1  {\ensuremath{\mathcal{C}_{#1}^{'\mathrm{(eff)}}}\xspace}       % 7
\def\Ope#1    {\ensuremath{\mathcal{O}_{#1}}\xspace}                       % 2
\def\Opep#1   {\ensuremath{\mathcal{O}_{#1}^{'}}\xspace}                    % 7

\newcommand{\tev}{\ifthenelse{\boolean{inbibliography}}{\ensuremath{~T\kern -0.05em eV}\xspace}{\ensuremath{\mathrm{\,Te\kern -0.1em V}}}\xspace}
\newcommand{\gev}{\ensuremath{\mathrm{\,Ge\kern -0.1em V}}\xspace}
\newcommand{\mev}{\ensuremath{\mathrm{\,Me\kern -0.1em V}}\xspace}
\newcommand{\kev}{\ensuremath{\mathrm{\,ke\kern -0.1em V}}\xspace}
\newcommand{\ev}{\ensuremath{\mathrm{\,e\kern -0.1em V}}\xspace}
\newcommand{\gevc}{\ensuremath{{\mathrm{\,Ge\kern -0.1em V\!/}c}}\xspace}
\newcommand{\mevc}{\ensuremath{{\mathrm{\,Me\kern -0.1em V\!/}c}}\xspace}
\newcommand{\gevcc}{\ensuremath{{\mathrm{\,Ge\kern -0.1em V\!/}c^2}}\xspace}
\newcommand{\gevgevcccc}{\ensuremath{{\mathrm{\,Ge\kern -0.1em V^2\!/}c^4}}\xspace}
\newcommand{\mevcc}{\ensuremath{{\mathrm{\,Me\kern -0.1em V\!/}c^2}}\xspace}

\def\mum  {\ensuremath{{\,\upmu\rm m}}\xspace}

\def\nb {\ensuremath{\rm \,nb}\xspace}

\def\gsim{{~\raise.15em\hbox{$>$}\kern-.85em
          \lower.35em\hbox{$\sim$}~}\xspace}
\def\lsim{{~\raise.15em\hbox{$<$}\kern-.85em
          \lower.35em\hbox{$\sim$}~}\xspace}

\def\sPlot{\mbox{\em sPlot}}

\def\pt         {\mbox{$p_{\rm T}$}\xspace}

\newcommand{\lum} {\ensuremath{\mathcal{L}}\xspace}
\def\evtgen     {\mbox{\textsc{EvtGen}}\xspace}

\def\geant      {\mbox{\textsc{Geant4}}\xspace}

\def\photos     {\mbox{\textsc{Photos}}\xspace}

\def\pythia     {\mbox{\textsc{Pythia}}\xspace}

\def\tell1  {TELL1\xspace}
\def\ukl1   {UKL1\xspace}

\newcommand{\eg}{\mbox{\itshape e.g.}\xspace}

\usepackage{cite} % Allows for ranges in citations
\usepackage{mciteplus}

\newcommand{\xx}{\ensuremath{\kern 0.5em }}

\newcommand{\OneSinpA}{\ensuremath{380\pm\, 35\pm\, 21\,{\rm \nb}}}
\newcommand{\OneSinAp}{\ensuremath{295\pm\, 56\pm\, 29\,{\rm \nb}}}
\newcommand{\OneSinpAc}{\ensuremath{211\pm\, 23\pm\, 11\,{\rm \nb}}}
\newcommand{\OneSinApc}{\ensuremath{282\pm\, 53\pm\, 25\,{\rm \nb}}}
\newcommand{\TwoSinpA}{\ensuremath{\xx75\pm\, 19\pm\, \xx5\,{\rm \nb}}}
\newcommand{\ThreeSinpA}{\ensuremath{\xx27\pm\, 16\pm\, \xx4\,{\rm \nb}}}
\newcommand{\TwoSinAp}{\ensuremath{\xx81\pm\, 39\pm\, 18\,{\rm \nb}}}
\newcommand{\ThreeSinAp}{\ensuremath{\xx\xx5\pm\, 26\pm\, \xx5\,{\rm \nb}}}

\def\pPb {\ensuremath{p\mathrm{Pb}}\xspace}
\def\pA {\ensuremath{p\mathrm{A}}\xspace}
\def\PbPb {\ensuremath{\mathrm{PbPb}}\xspace}
\def\sPlot{\mbox{\em sPlot}\xspace}

\def\YnS{\ensuremath{\PUpsilon{(nS)}}\xspace}% no space before {...}!
\def\nS  {{\YnS}}

\usepackage{multirow} % for complicated table
\usepackage{booktabs} % for complicated table
\usepackage{rotating}

\begin{document}

%%%%%%%%%%%%%%%%%%%%%%%%%
%%%%% Title     %%%%%%%%%
%%%%%%%%%%%%%%%%%%%%%%%%%
\renewcommand{\thefootnote}{\fnsymbol{footnote}}
\setcounter{footnote}{1}

% %%%%%%% CHOOSE TITLE PAGE--------
%\onecolumn
%%%%%%%%%%%%%%%%%%%%%%%%%
%%%%%  TITLE PAGE  %%%%%%
%%%%%%%%%%%%%%%%%%%%%%%%%
\begin{titlepage}
\pagenumbering{roman}

% Header ---------------------------------------------------
\vspace*{-1.5cm}
\centerline{\large EUROPEAN ORGANIZATION FOR NUCLEAR RESEARCH (CERN)}
\vspace*{1.5cm}
\hspace*{-0.5cm}
\begin{tabular*}{\linewidth}{lc@{\extracolsep{\fill}}r}
\ifthenelse{\boolean{pdflatex}}% Logo format choice
{\vspace*{-2.7cm}\mbox{\!\!\!\includegraphics[width=.14\textwidth]{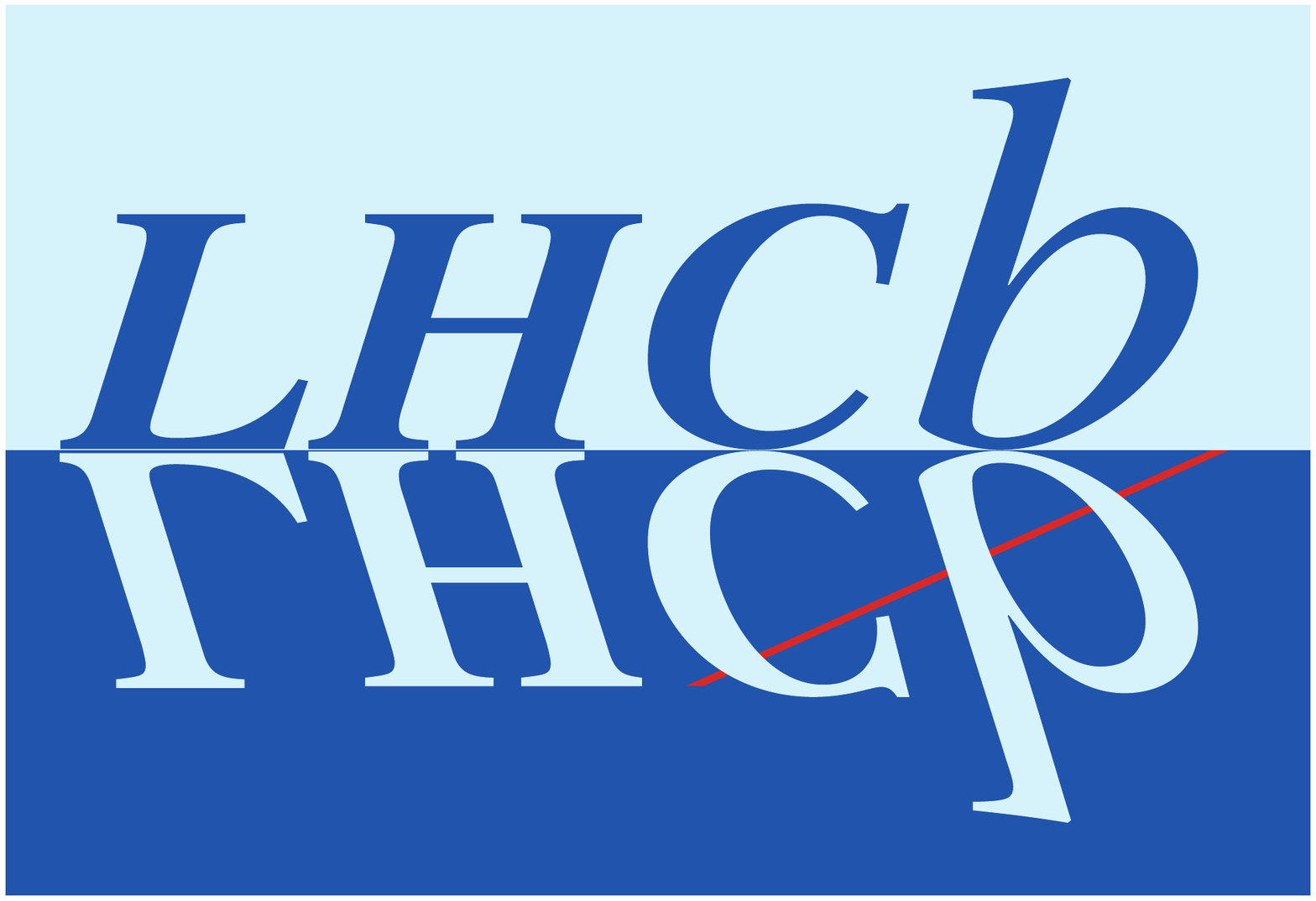}} & &}%
{\vspace*{-1.2cm}\mbox{\!\!\!\includegraphics[width=.12\textwidth]{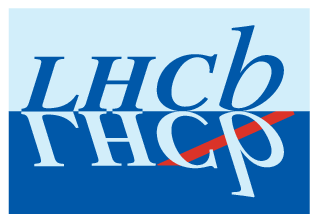}} & &}%
\\
 & & CERN-PH-EP-2014-102 \\  % ID 
 & & LHCb-PAPER-2014-015 \\  % ID 
% & & \today \\ % Date - Can also hardwire e.g.: 23 March 2010
 %& & 20 May 2014 \\
 & & 15 October 2014 \\
 & & \\
% not in paper \hline
\end{tabular*}

\vspace*{1.5cm}

% Title --------------------------------------------------
{\bf\boldmath\huge
\begin{center}
Study of $\PUpsilon$ production and cold nuclear matter effects in $\pPb$ collisions
     % at $5\tev$
      at $\sqrt{s_{\mbox{\small{\bf{\it NN}}}}}=5\mathrm{\,Te\kern -0.1em V}$
\end{center}
}

\vspace*{1.0cm}

% Authors -------------------------------------------------
\begin{center}
The LHCb collaboration\footnote{Authors are listed on the following pages.}
\end{center}

\vspace{\fill}

% Abstract -----------------------------------------------
\begin{abstract}
\noindent
Production of $\PUpsilon$\/ mesons in proton-lead 
collisions at a nucleon-nucleon centre-of-mass energy
$\sqrt{s_{\mbox{\tiny{\it NN}}}}=5\mathrm{\,Te\kern -0.1em V}$\/
is studied with the {\mbox{LHCb}\xspace}\ detector. The analysis 
is based on a data sample corresponding to an integrated luminosity of
$1.6~\mathrm{nb}^{-1}$. The $\PUpsilon$\/ mesons of transverse momenta up 
to $15\gevc$ are reconstructed in the dimuon decay mode.
The rapidity coverage in the centre-of-mass system is $1.5<y<4.0$\/ (forward region)
and $-5.0<y<-2.5$ (backward region). 
The forward-backward production ratio and the nuclear modification 
factor for $\OneS$\/ mesons are determined. 
The data are compatible with the predictions for a suppression of $\OneS$ production 
with respect to proton-proton collisions in the forward region, 
and an enhancement in the backward region.
The suppression is found to be smaller than in the case of prompt $\jpsi$ mesons.
\end{abstract}

\vspace*{1.0cm}

\begin{center}
  published in JHEP 
\end{center}

\vspace{\fill}

{\footnotesize 
\centerline{\copyright~CERN on behalf of the \lhcb collaboration, license \href{http://creativecommons.org/licenses/by/3.0/}{CC-BY-3.0}.}}
\vspace*{2mm}

\end{titlepage}

%%%%%%%%%%%%%%%%%%%%%%%%%%%%%%%%
%%%%%  EOD OF TITLE PAGE  %%%%%%
%%%%%%%%%%%%%%%%%%%%%%%%%%%%%%%%

%  empty page follows the title page ----
\newpage
\setcounter{page}{2}
\mbox{~}
\newpage

% Author List ----------------------------
%  You need to get a new author list!
%%%%%%%%%%%%%%%%%%%%%%%%%%%%%%%%%%%%%%%%%%
\centerline{\large\bf LHCb collaboration}
\begin{flushleft}
\small
R.~Aaij$^{41}$, 
B.~Adeva$^{37}$, 
M.~Adinolfi$^{46}$, 
A.~Affolder$^{52}$, 
Z.~Ajaltouni$^{5}$, 
J.~Albrecht$^{9}$, 
F.~Alessio$^{38}$, 
M.~Alexander$^{51}$, 
S.~Ali$^{41}$, 
G.~Alkhazov$^{30}$
% ...

{\footnotesize \it
$ ^{1}$Centro Brasileiro de Pesquisas F\'{i}sicas (CBPF), Rio de Janeiro, Brazil\\
% ...
$ ^{5}$Clermont Universit\'{e}, Universit\'{e} Blaise Pascal, CNRS/IN2P3, LPC, Clermont-Ferrand, France\\
% ...
$ ^{9}$Fakult\"{a}t Physik, Technische Universit\"{a}t Dortmund, Dortmund, Germany\\
% ...
$ ^{30}$Petersburg Nuclear Physics Institute (PNPI), Gatchina, Russia\\
% ...
$ ^{37}$Universidad de Santiago de Compostela, Santiago de Compostela, Spain\\
$ ^{38}$European Organization for Nuclear Research (CERN), Geneva, Switzerland\\
% ...
$ ^{41}$Nikhef National Institute for Subatomic Physics, Amsterdam, The Netherlands\\
% ...
$ ^{46}$H.H. Wills Physics Laboratory, University of Bristol, Bristol, United Kingdom\\
% ...
$ ^{51}$School of Physics and Astronomy, University of Glasgow, Glasgow, United Kingdom\\
$ ^{52}$Oliver Lodge Laboratory, University of Liverpool, Liverpool, United Kingdom\\
% ...
}
\end{flushleft}
%%%%%%%%%%%%%%%%%%%%%%%%%%%%%%%%%%%%%%%%%%

\cleardoublepage

%\twocolumn
% %%%%%%%%%%%%% ---------

\renewcommand{\thefootnote}{\arabic{footnote}}
\setcounter{footnote}{0}

%%%%%%%%%%%%%%%%%%%%%%%%%%%%%%%%
%%%%%  Table of Content   %%%%%%
%%%%%%%%%%%%%%%%%%%%%%%%%%%%%%%%
%%%% Uncomment next 2 lines if desired
%\tableofcontents
%\cleardoublepage

%%%%%%%%%%%%%%%%%%%%%%%%%
%%%%% Main text %%%%%%%%%
%%%%%%%%%%%%%%%%%%%%%%%%%

\pagestyle{plain} % restore page numbers for the main text
\setcounter{page}{1}
\pagenumbering{arabic}

%% Uncomment during review phase. 
%% Comment before a final submission.
%\linenumbers

% You can include short sections directly in the main tex file.
% However, for larger papers it is desirable to split the text into
% several semiautonomous files, which can be revised independently.
% This is especially useful when developing a document in
% collaboration with several people, since then different parts can be
% edited independently.  This type of file organization is shown here.
% 
\section{Introduction}
\label{sec:Introduction}
\noindent 
Heavy quarkonia are produced at the early stage of ultra-relativistic
heavy-ion collisions and probe
the existence of the quark-gluon plasma (QGP), a hot and dense nuclear medium. 
Due to colour screening effects in the QGP, the yield of 
heavy quarkonia in heavy-ion collisions is expected to be suppressed 
with respect to proton-proton ($pp$) collisions~\cite{Matsui:1986dk}.
Heavy quarkonium production can also be suppressed by
normal nuclear matter effects, often referred to as
cold nuclear matter (CNM) effects, 
such as nuclear shadowing (antishadowing) effects, 
energy loss of the heavy quark or the heavy quark pair in the medium or nuclear absorption. 
Shadowing and antishadowing effects
\cite{Ferreiro:2013pua,Albacete:2013ei,Adeluyi:2013tuu,Chirilli:2012jd,Ferreiro:2011xy}
describe how the parton densities are modified when a nucleon is bound inside a nucleus.
%A coherent treatment of energy loss for the inital state partons and final state $Q\bar{Q}$ pairs
%in nuclear matter is described in Refs.~\cite{Arleo:2012rs,Arleo:2013zua},
%where $Q$ presents a charm or bottom quark. 
A coherent treatment of energy loss for the inital state partons and final state $c\bar{c}$ or $b\bar{b}$ pairs 
in nuclear matter is described in Refs.~\cite{Arleo:2012rs,Arleo:2013zua}.
Nuclear absorption is a final-state effect caused by the break-up of these pairs
due to the inelastic scattering with the nucleons.
The importance of studying absorption effects for quarkonia 
in the high energy heavy-ion proton collisions
is discussed in Refs.~\cite{Kharzeev:1995br,Kharzeev:1995id,Kharzeev:1996yx}, 
and the energy and rapidity dependence was studied in Ref.~\cite{Lourenco:2008sk}.
%which are present regardless of whether a QGP is created or not. 
%The CNM effects influencing quarkonium production include: 
%initial-state nuclear effects on the parton densities (shadowing or antishadowing);
%initial-state parton energy loss and final-state energy loss;
%final-state absorption by nucleons, expected to be negligible 
%at the LHC energies; and intrinsic heavy flavour content in the nucleons 
%\cite{Ferreiro:2013pua,Arleo:2012rs,Albacete:2013ei,Adeluyi:2013tuu,
%Chirilli:2012jd,Chirilli:2012sk,Arleo:2013zua,Ferreiro:2011xy}. 
The main models describing quarkonium production in hadron collisions 
are the colour-singlet model (CSM) \cite{Chang:1979nn,Baier:1981uk,Campbell:2007ws,Artoisenet:2008fc},
the colour-evaporation model (CEM) \cite{CEM} 
and non-relativistic quantum chromodynamics (NRQCD) \cite{Bodwin:1994jh,Bodwin:1994jh:Erratum,Cho:1995vh,Cho:1995ce}.

The $\pA$ collisions in which a QGP is not expected to be created, 
provide a unique opportunity to study CNM effects and 
to constrain the nuclear parton distribution functions describing
the partonic structure of matter. 
These measurements offer crucial information to disentangle CNM effects from the effects of QGP in nucleus-nucleus collisions. 
%A lot of endeavours were made to investigate the CNM effects on charmonia 
Several measurements of CNM effects were performed
by the fixed-target experiments 
at the SPS~\cite{Abreu:1998ee,Alessandro:2006jt,Abreu:1998rx,Arnaldi:2010ky}, 
Fermilab~\cite{Leitch:1999ea} 
and DESY~\cite{Abt:2008ya}. 
%Efforts are also made by deuteron-gold collisions at RHIC~\cite{}.
With the proton-lead (\pPb) data collected in 2013, CNM effects have been studied by
the \lhcb experiment with measurements of the differential production cross-sections of prompt \jpsi mesons
and \jpsi from $b$-hadron decays~\cite{LHCb-PAPER-2013-052}, 
and by the ALICE experiment using measurements of inclusive \jpsi production~\cite{ALICE-Jpsi-in-pA}.
Unambiguous CNM effects have been observed, in agreement with theoretical predictions.

The study of bottomonia, $\OneS$, $\TwoS$\/ and $\ThreeS$ mesons,
denoted generically by $\PUpsilon$\/ in the following, provides 
complementary information about CNM effects  to that from \jpsi production.
For example, the $\OneS$\/ meson 
can survive in the QGP at higher temperatures than other heavy quarkonia owing to its higher 
binding energy \cite{Karsch:1990wi,Digal:2001ue}. 
As a consequence, 
based on the prediction that the dissociation of $\PUpsilon$\/
states in the QGP occurs sequentially according to their different
binding energies~\cite{Digal:2001ue},
it is interesting to determine the production ratios of excited $\PUpsilon$ mesons,
\begin{equation}
\label{equ:R_ns}
R^{nS/1S}\equiv
\frac{\sigma(\nS)\times\BF(\nS\to\mumu)}
{\sigma(\OneS)\times\BF(\OneS\to\mumu)},\quad n=2,3,
\end{equation}
where $\sigma$ represents the cross-section for the production
of the indicated meson and $\BF$ represents the branching fraction for its dimuon decay mode.
The production ratios $R^{nS/1S}$ have been measured 
in \pPb \cite{Chatrchyan:2013nza} and $\PbPb$ \cite{Chatrchyan:2012lxa} collisions for central rapidities 
by the CMS experiment and the ratios of these quantities to $R^{nS/1S}$ measured
in $pp$ collisions show clear sequential suppression of $\PUpsilon$ production,
which indicates stronger (cold or hot) nuclear matter effects on the excited $\PUpsilon$ states.
\lhcb can extend those studies to the forward and backward rapidity regions.
From the theoretical point of view, 
predictions for bottomonia are more reliable than those for charmonia 
owing to the heavier quark masses and lower quark velocities.

In this analysis, the inclusive production cross-sections of $\PUpsilon$\/ mesons
are measured in \pPb collisions at a nucleon-nucleon centre-of-mass energy 
$\sqrt{s_{\mbox{\tiny{\it NN}}}}=5\tev$ at \lhcb. 
Based on the cross-section measurements,
the production ratios $R^{nS/1S}$ are evaluated
 and the CNM effects for $\OneS$\/ mesons are studied.
The LHCb detector is a single-arm forward spectrometer~\cite{Alves:2008zz} that covers 
the \mbox{pseudorapidity} region $2<\eta<5$ in $pp$ collisions. 
To allow for measurements of \pA collisions at both positive and negative rapidity,
where rapidity is defined with respect to the direction of the proton, 
the proton and lead beams were interchanged approximately halfway during the \pPb data taking period.
Owing to the asymmetry in the energy per nucleon in the two beams, the nucleon-nucleon centre-of-mass 
system has a rapidity of $+0.465~(-0.465)$\/ in the laboratory frame for 
the forward (backward) collisions,
where forward (backward) is defined as positive (negative) rapidity.
For the measurements described here rapidity ranges of $1.5<y<4.0$\/ and $-5.0<y<-2.5$ are studied.

\section{Detector and data set}
\label{sec:Detector}
The \mbox{LHCb}\xspace detector~\cite{Alves:2008zz} is 
designed for the study of particles containing $b$ or $c$ 
quarks. The detector includes a high-precision tracking system
consisting of a silicon-strip vertex detector ({VELO\xspace}) %~\cite{LHCb-DP-2010-001}
surrounding the interaction region, a large-area silicon-strip detector 
located upstream of a dipole magnet with a bending power of about
$4{\rm\,Tm}$, and three stations of silicon-strip detectors and straw
drift tubes~\cite{LHCb-DP-2013-003} placed downstream of the magnet. 
The combined tracking system provides a momentum resolution
with a relative uncertainty that varies from 0.4\% at low momentum to 0.6\% at 100\gevc,
and an impact parameter measurement with a resolution of 20\mum for
charged particles with large transverse momentum, \pt. 
Different types of charged hadrons are distinguished using information
from two ring-imaging Cherenkov (RICH) detectors~\cite{LHCb-DP-2012-003}. 
Photon, electron and hadron candidates are identified by a calorimeter system consisting of
scintillating-pad and preshower detectors, an electromagnetic
calorimeter and a hadronic calorimeter. Muons are identified by a
system composed of alternating layers of iron and multiwire
proportional chambers~\cite{LHCb-DP-2012-002}.
The trigger~\cite{LHCb-DP-2012-004} consists of a
hardware stage, based on information from the calorimeter and muon
systems, followed by a software stage, which applies a full event
reconstruction.

The data sample used for this analysis was acquired during the 
$p\mathrm{Pb}$\/ run in early 2013 and corresponds to an integrated 
luminosity of $1.1~\mathrm{nb}^{-1}$ ($0.5~\mathrm{nb}^{-1}$)
for forward (backward) collisions. 
The hardware trigger was employed as an interaction trigger
that rejected empty events. The software trigger required 
one well-reconstructed charged particle with hits in the muon system and a transverse 
momentum greater than $600\mathrm{\,Me\kern -0.1em V\!/}c$.

Simulated samples based on $pp$\/ collisions at $8\mathrm{\,Te\kern -0.1em V}$\/ 
are reweighted according to the track multiplicity 
to reproduce the experimental data at $5\mathrm{\,Te\kern -0.1em V}$.
The effect of the asymmetric beam energies in \pPb collisions and different detector occupancies 
have been taken into account for the determination of the efficiencies.
In the simulation, $pp$ collisions are generated using \pythia~6.4
\cite{Sjostrand:2006za} with a specific \lhcb\ configuration
\cite{LHCb-PROC-2010-056}. Hadron decays are described by \evtgen\
\cite{Lange:2001uf}, where final-state radiation is generated using 
\photos \cite{Golonka:2005pn}. The interactions of the generated particles 
with the detector and its response are implemented using the \geant\
toolkit \cite{Allison:2006ve, *Agostinelli:2002hh} as described in
Ref.~\cite{LHCb-PROC-2011-006}.

\section{Cross-section determination}
\label{sec:UpsilonEventSelectionXsection}
The total cross-section is measured for 
$\OneS$, $\TwoS$\/ and $\ThreeS$\/ mesons in the kinematic region $\pt<15\gevc$\/ 
and $1.5<y<4.0$\/ ($-5.0<y<-2.5$) for the forward (backward) sample.
The cross-section is also measured in the common rapidity coverage of the forward and backward samples, 
$2.5<|y|<4.0$, to study CNM effects.
The product of the total production cross-sections and the branching fractions for $\PUpsilon(nS)$\/ mesons
is given by
\begin{equation}
\label{eq:UpsilonDoubleDifferential}
  \sigma(\nS)\times\BF(\nS\to\mup\mun) 
 =\frac{N^{\mathrm{cor}}(\nS\to\mup\mun)}{\lum}, \quad n=1,2,3,
\end{equation}
where
$N^\mathrm{cor}(\nS\to\mup\mun)$\/ is the efficiency-corrected number 
of signal candidates reconstructed with dimuon final states in the given 
$\pt$\/ and $y$\/ region, and \lum is the integrated luminosity, calibrated 
by means of van der Meer scans \cite{vanderMeer:1968zz,LHCb-PAPER-2013-052} for each beam configuration separately.

The strategy for the $\PUpsilon$ cross-section measurement 
follows Refs.~\cite{LHCb-PAPER-2011-036,LHCb-PAPER-2013-016,LHCb-PAPER-2013-066}.
The $\PUpsilon$\ candidates are reconstructed from two oppositely charged particles 
consistent with a muon hypothesis based on particle identification information 
from the RICH detectors, the calorimeters and the muon system. 
Each particle must have a $\pt$\/ above 
$1~\gevc$\/ and a good track fit quality. 
The two muon candidates are required to originate from a common vertex.

An unbinned extended maximum likelihood fit to the invariant mass distribution 
of the selected candidates is performed to determine the signal yields 
of $\OneS$, $\TwoS$\/ and $\ThreeS$\/ mesons
in a fit range $8400<m_{\mumu}<11400\mevcc$.
To describe the $\OneS$, $\TwoS$\/ and $\ThreeS$\/ signal components, 
a sum of three Crystal Ball (CB) functions~\cite{Skwarnicki:1986xj} is used, 
while the combinatorial background is modelled with an exponential function.

The shape parameters of the CB functions have been fixed using large samples collected 
in $pp$ collisions~\cite{LHCb-PAPER-2013-016},
which determine the mass resolution for the $\OneS$\/ to be $43.0\mevcc$.
The resolutions for the $\TwoS$ and $\ThreeS$ signals are obtained 
by scaling this value by the ratio of their masses to the $\OneS$ meson mass \cite{PDG2012}. 

\begin{figure}[tb]
\begin{center}
\includegraphics[width=0.49 \textwidth]{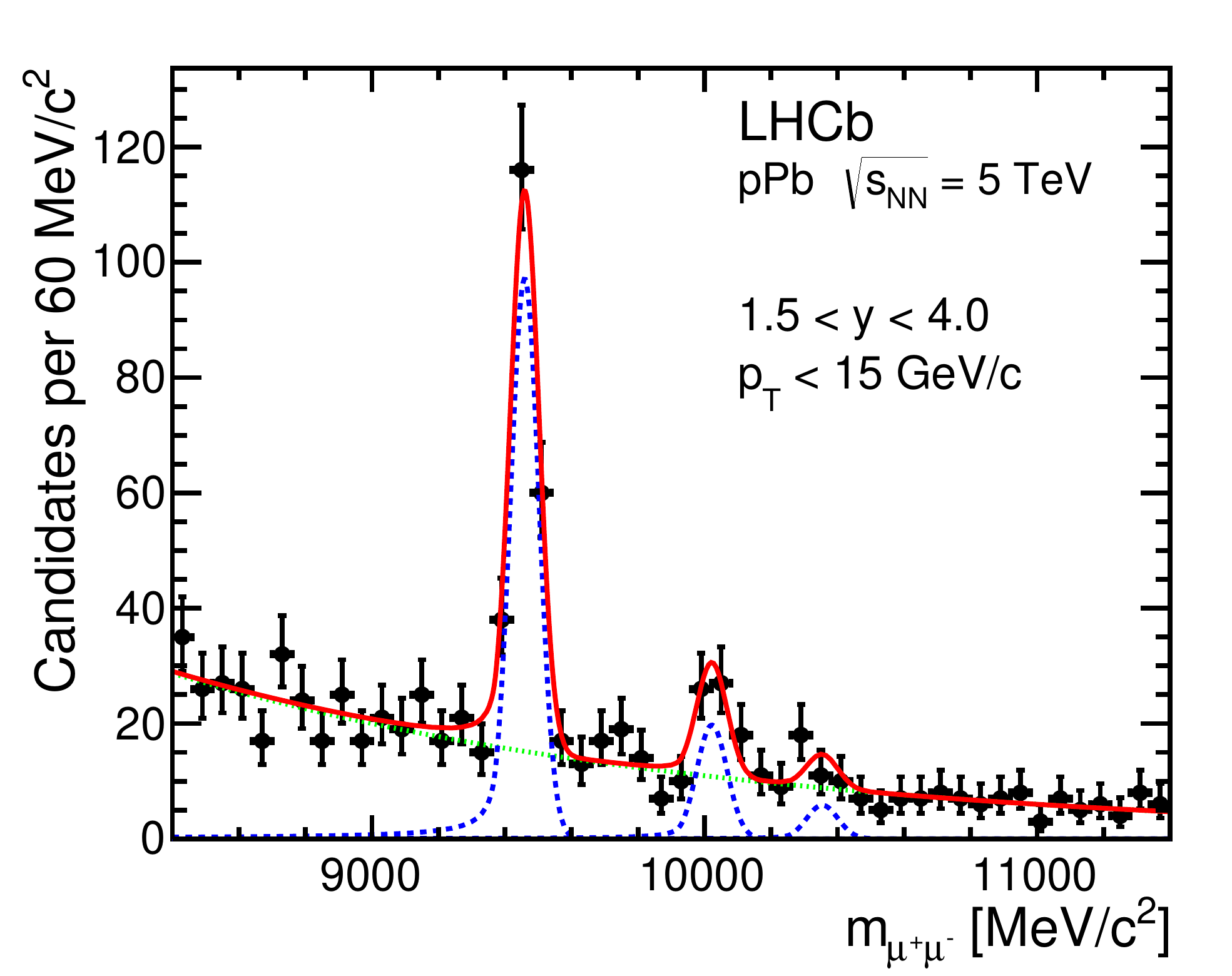}
\includegraphics[width=0.49 \textwidth]{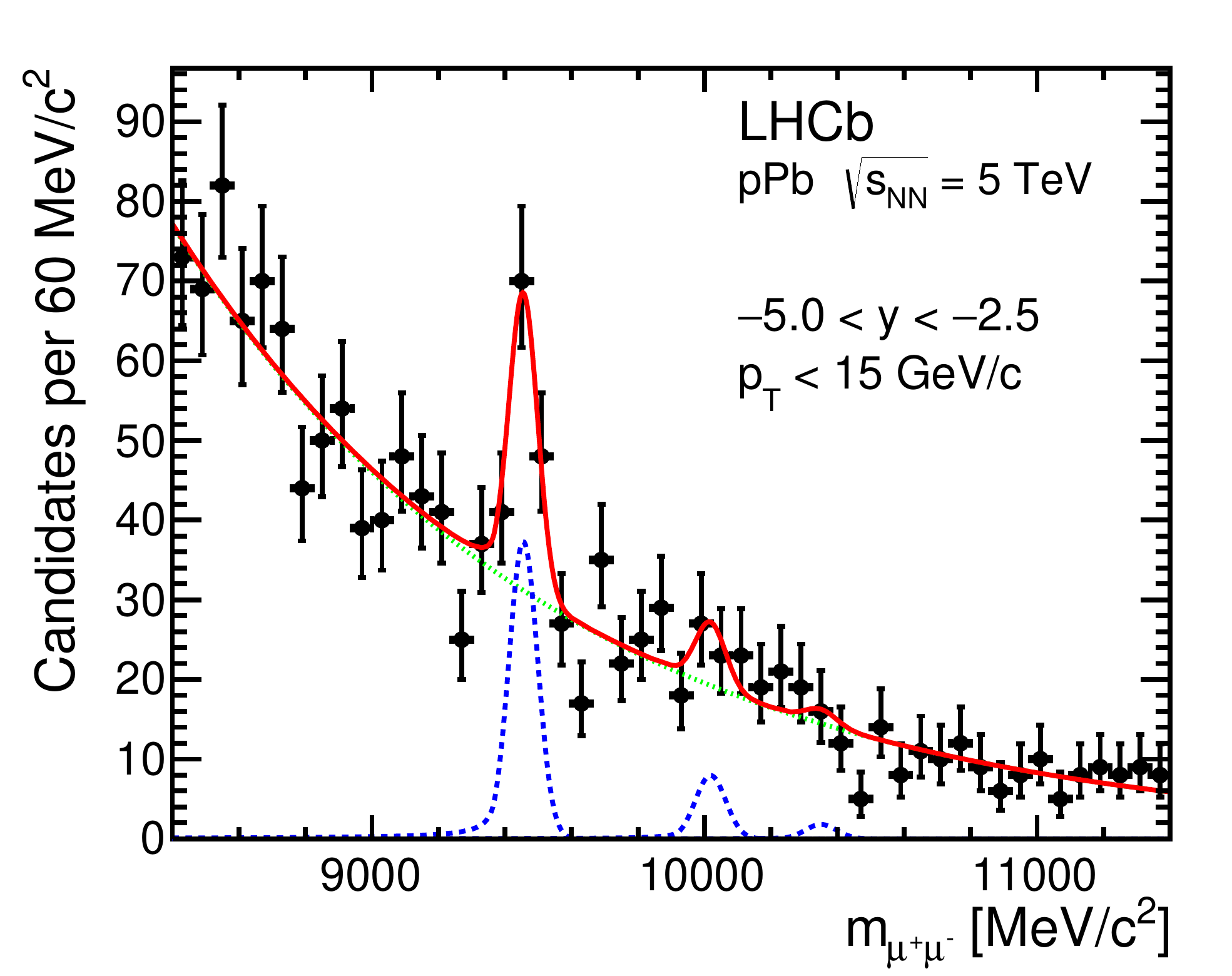}
\vspace*{-0.5cm}
\end{center}
\caption{ \small 
        Invariant mass distribution of $\mu^{+}\mu^{-}$ pairs 
        in the (left) forward 
        and (right) backward samples of $\pPb$ collisions.
        The transverse momentum range is $\pt<15\gevc$.
        The rapidity range is $1.5<y<4.0$ ($-5.0<y<-2.5$) 
        for the forward (backward) sample.
        The black dots are the data points, 
        the blue dashed curve indicates the signal component,
        the green dotted curve represents the combinatorial background,
        and the red solid curve is the sum of the signal and background components.
        }
\label{fig:UpsilonMass}
\end{figure}

Figure~\ref{fig:UpsilonMass} shows the dimuon invariant mass distributions in the $p\mathrm{Pb}$\/ 
forward and backward samples, with the fit results superimposed.
In the backward sample higher combinatorial background is observed due to the larger track multiplicity.
The signal yields obtained from the fit are $N_{\OneS} = 189\pm16$ ($72\pm 14$), 
$N_{\TwoS} = 41\pm 9$ ($17\pm 10$), and $N_{\ThreeS} = 13\pm7$ ($4\pm 8$) in the forward (backward) sample.
The yields of $\OneS$\/ mesons with $2.5<|y|<4.0$
are $122\pm13$ in the forward sample and $70\pm13$ in the backward sample.
The uncertainties are statistical only.

A signal weight factor, $\omega_i$,
is assigned to each candidate using the \sPlot technique~\cite{Pivk:2004ty}
with the dimuon invariant mass as the discriminating variable. 
The efficiency-corrected signal yield $N^{\mathrm{cor}}$\/ is then calculated 
through an event-by-event efficiency correction $\epsilon_{i}$ as
\begin{equation}
\label{equ:event-by-event}
N^{\mathrm{cor}}=\sum_{i}{\omega_{i}/\epsilon_{i}},
\end{equation}
where the sum runs over all events.
The total signal efficiency, which depends on the $\pt$ and $y$ of the $\PUpsilon$ mesons,
is the product of the geometric acceptance, reconstruction and selection, 
muon identification, and trigger efficiencies.
The product of the acceptance, reconstruction and selection efficiencies is determined in
fine \pt and $y$ bins with simulated samples.
The simulated events are
reweighted according to the track multiplicity observed in data and
corrected to account for small differences in the
track-reconstruction efficiency between data and simulation~\cite{Jaeger:1402577,Archilli:2013npa}. 
In the selected rapidity range the reconstruction and selection efficiency varies between 30\% and 81\%.
The muon identification efficiency
is obtained as a function of momentum and transverse momentum
by a data-driven tag-and-probe approach using a $\jpsi\to\mu^{+}\mu^{-}$ sample~\cite{Jaeger:1402577}.
For $\PUpsilon$ candidates this efficiency is generally larger than 90\%.
%The trigger efficiency is determined from the data using a sample of events containing 
%$\OneS$\/ mesons triggered independently of the signal, and is around 95\%.
The trigger efficiency was determined using a sample of $\OneS$ decays into muon pairs
that did not require the muons to be in the trigger, and is around 95\%. 
The corresponding uncertainty is described in the following section.
Here the much more abundant \jpsi decays are not used since the trigger efficiency 
%is sensitive to the transverse momentum spectrum of the muon. 
depends on the muon transverse momentum.
%Nevertheless, the final uncertainty of the trigger efficiency of 2.1\% due to the small number of events
%in this study is still much smaller than the statistical uncertainties of the measurements.

%%%%%%%%%%%%%%%%%%%%%%%%%%%%%%%%%%%%%%%%%%%%%%%%%%%%%%%
\section{Systematic uncertainties}
\label{sec:systematics}
The systematic uncertainties of this analysis
are summarised in Table \ref{tab:UpsilonSystematics}.
They are added in quadrature to obtain the total systematic uncertainty.

Due to the finite size of the $\jpsi$ calibration sample, the systematic
uncertainty of the muon identification efficiency obtained from the 
tag-and-probe approach
is $1.3\%$. 
The uncertainty due to the track reconstruction efficiency is estimated to be $1.5\%$ 
by varying within its uncertainty the correction applied to 
the muon reconstruction efficiency.

The systematic uncertainty due to the choice of the fit model used to 
describe the shape of the dimuon mass distribution is estimated
by varying the fixed parameters of the CB function, or by using a polynomial 
function, whose parameters are determined by the fit, to describe the background shape. 
The largest difference in yields of each resonance with respect to the nominal result
is considered as the systematic uncertainty.

%%%%%%%%%%%%%%%%%%%%%%%%%%%%%%%%%%%%%%%%%%%%%%%
\begin{table}[tp]
\caption{\small 
Relative systematic uncertainties on the cross-sections, in percent, in the full rapidity range. 
The values in parenthesis refer specifically to $\OneS$ measurements when systematic uncertainties
in the common rapidity range $2.5<|y|<4.0$ are notably different.
}
\begin{center}
\begin{tabular}{lccc|ccc}
\toprule
                            & \multicolumn{3}{c|}{Forward}  &  \multicolumn{3}{c}{Backward} \\
Source                      & \OneS & \TwoS & \ThreeS & \OneS & \TwoS & \ThreeS \\
\hline 
Muon identification         &  1.3 &  1.3 &  1.3 &  1.3 &  1.3 &  1.3 \\
Tracking efficiency         &  1.5 &  1.5 &  1.5 &  1.5 &  1.5 &  1.5 \\
Mass fit model              & 1.1 (1.0) & 4.9 & 13 & 1.8 (1.7)&19 & 90 \\
Luminosity                  &  1.9 &  1.9 &  1.9 &  2.1 &  2.1 &  2.1 \\
Trigger                     &  2.1 &  2.1 &  2.1 &  5.0 &  5.0 &  5.0 \\
MC generation kinematics    & 3.9 (3.8) & 3.9 & 3.9 & 7.6 (6.3)&7.6 &7.6 \\ 
Reconstruction              &  1.5 &  1.5 &  1.5 &  1.5 &  1.5 &  1.5 \\
\hline 
Total                       & 5.5 (5.4) & 7.3 & 14 & 9.8 (8.8) & 21 & 91 \\
\bottomrule
\end{tabular}\end{center}
\label{tab:UpsilonSystematics}
\end{table}
%%%%%%%%%%%%%%%%%%%%%%%%%%%%%%%%%%%%%%%%%%%%%%%

The luminosity is determined with an uncertainty of $1.9\%$ ($2.1\%$)
for the $\pPb$ forward (backward) sample from the rate of 
interactions that yield at least one reconstructed track
in the VELO. The absolute calibration is determined with van der Meer
scans, as described in Ref.~\cite{LHCb-PAPER-2013-052}.

The trigger efficiency in the forward sample is determined directly 
from the data using a sample
unbiased by the trigger decision.
The corresponding uncertainty is $2.1\%$.
Due to the limited sample size, the trigger efficiency in the backward sample
is estimated using the forward sample,
since it has been observed that the dependence of trigger efficiencies 
on the charged-particle multiplicity is small~\cite{LHCb-PAPER-2013-052}.
The systematic uncertainty is $5.0\%$,
taking into account the difference between the trigger efficiencies obtained using 
the forward and backward samples.

An uncertainty is introduced by the possible difference between the data and simulation samples of the $\pt$\/
and $y$\/ spectra inside each bin.
This is estimated 
by doubling the number of $\pt$\/ or $y$\/ bins in the efficiency
tables based on the simulated samples. In the forward (backward) 
sample, the difference to the nominal binning is $3.9\%$ ($7.6\%$) 
in the full rapidity range, and $3.8\%$ ($6.3\%$) in the common 
rapidity coverage. These differences are taken as systematic 
uncertainties.

The systematic uncertainties due to 
reconstruction effects, \eg track and vertexing quality,
have been studied in the $\jpsi$\/ analysis in $\pPb$\/ collisions~\cite{LHCb-PAPER-2013-052} 
and determined to be $1.5\%$.

Although the initial polarisation of the vector meson affects the efficiency,
 recent results show that the polarisations of the \OneS, \TwoS\ and \ThreeS\ mesons 
 are small in $pp$ collisions~\cite{Chatrchyan:2012woa}. 
In this analysis, we take them to be zero
and do not assign any systematic uncertainty to account for this assumption.

%%%%%%%%%%%%%%%%%%%%%%%%%%%%%%%%%%
\section{Results}
The products of production cross-sections and branching fractions for $\PUpsilon$ mesons with $\pt<15\gevc$
are measured for the different rapidity ranges to be
\begin{equation*}
\begin{split}
\sigma(\OneS,-5.0<y<-2.5)\times\BF(1S) &=\OneSinAp,\\
\sigma(\TwoS,-5.0<y<-2.5)\times\BF(2S) &=\TwoSinAp,\\
\sigma(\ThreeS,-5.0<y<-2.5)\times\BF(3S)& =\ThreeSinAp,\\
\sigma(\OneS,\ \xx1.5<y<\ \xx4.0)\times\BF(1S) &=\OneSinpA,\\
\sigma(\TwoS,\ \xx1.5<y<\ \xx4.0)\times\BF(2S) &=\TwoSinpA,\\
\sigma(\ThreeS,\ \xx1.5<y<\ \xx4.0)\times\BF(3S)& =\ThreeSinpA,
\end{split}
\end{equation*}
where the first uncertainty is statistical and the second systematic,
a convention also used in the following.
The variation in relative size of the statistical uncertainty compared 
to the signal yields is due to the variation of the event-by-event 
efficiencies and the variation of the signal-to-background ratio over the accessible phase space. 
In the common rapidity range $2.5<|y|<4.0$, the results for $\OneS$ production are 
\begin{equation*}
\begin{split}
&\sigma(\OneS,-4.0<y<-2.5)\times\BF(1S) =\OneSinApc,\\
&\sigma(\OneS,\ \xx2.5<y<\ \xx4.0)\times\BF(1S) =\OneSinpAc.
\end{split}
\end{equation*}

Using the results described above, the production ratios $R^{nS/1S}$ 
are measured to be
\begin{equation*}
\begin{split}
R^{2S/1S}(-5.0<y<-2.5)&=0.28\pm0.14\pm0.05,\\
R^{3S/1S}(-5.0<y<-2.5)&=0.02\pm0.09\pm0.02, \\
R^{2S/1S}(\ \xx1.5<y<\ \xx4.0)&=0.20\pm0.05\pm0.01,\\
R^{3S/1S}(\ \xx1.5<y<\ \xx4.0)&=0.07\pm0.04\pm0.01.
\end{split}
\end{equation*}
In these ratios all the systematic uncertainties cancel except for 
those due to the mass fit model. 
The measurements of $R^{nS/1S}$ in \pPb collisions are compatible with those in $pp$\/ collisions
\cite{LHCb-PAPER-2011-036,LHCb-PAPER-2013-016,LHCb-PAPER-2013-066}.

\begin{figure}[ht]
  \begin{center}
    \vspace*{-0.5cm}
    \includegraphics[scale=0.35]{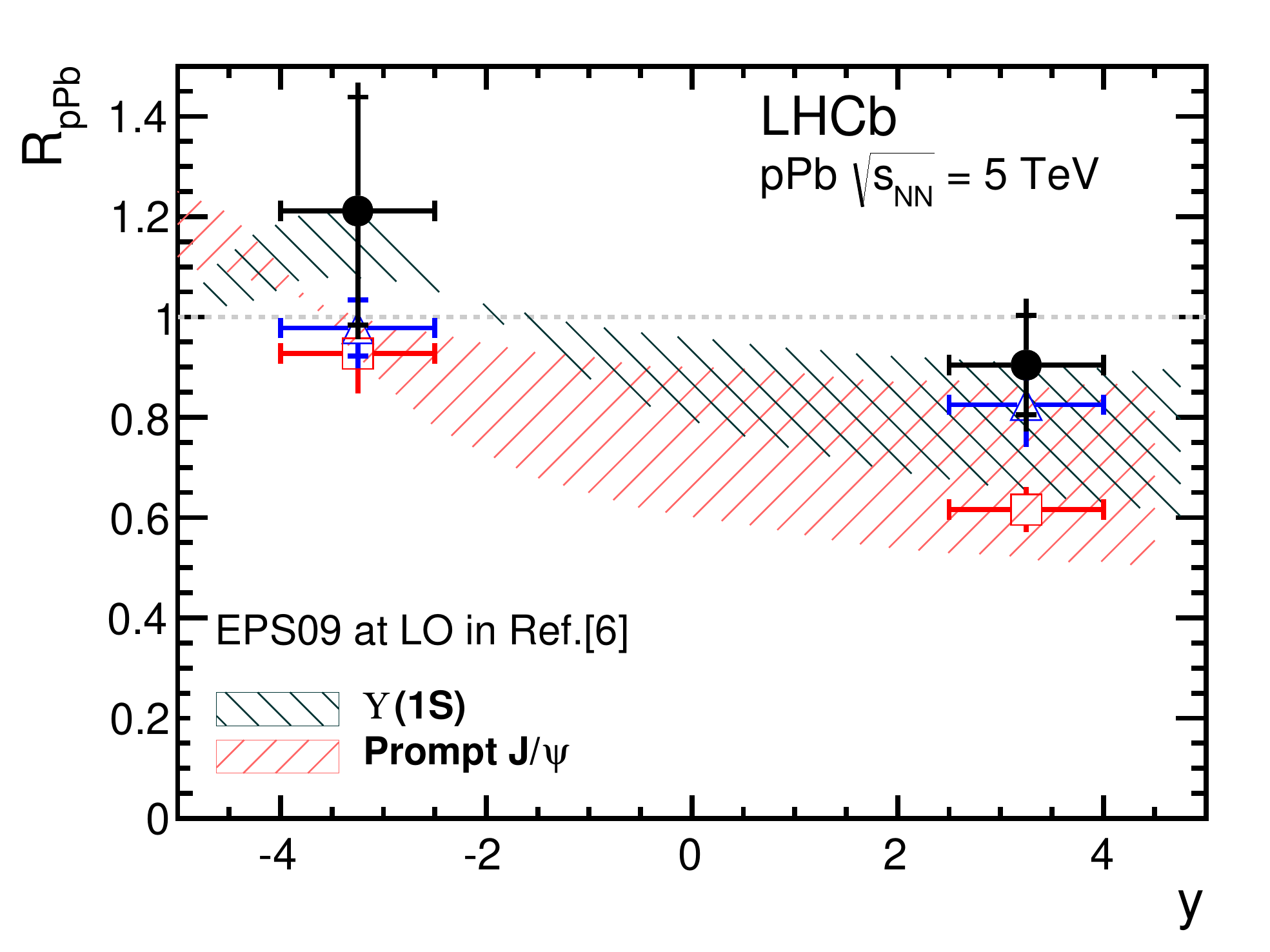}
    \includegraphics[scale=0.35]{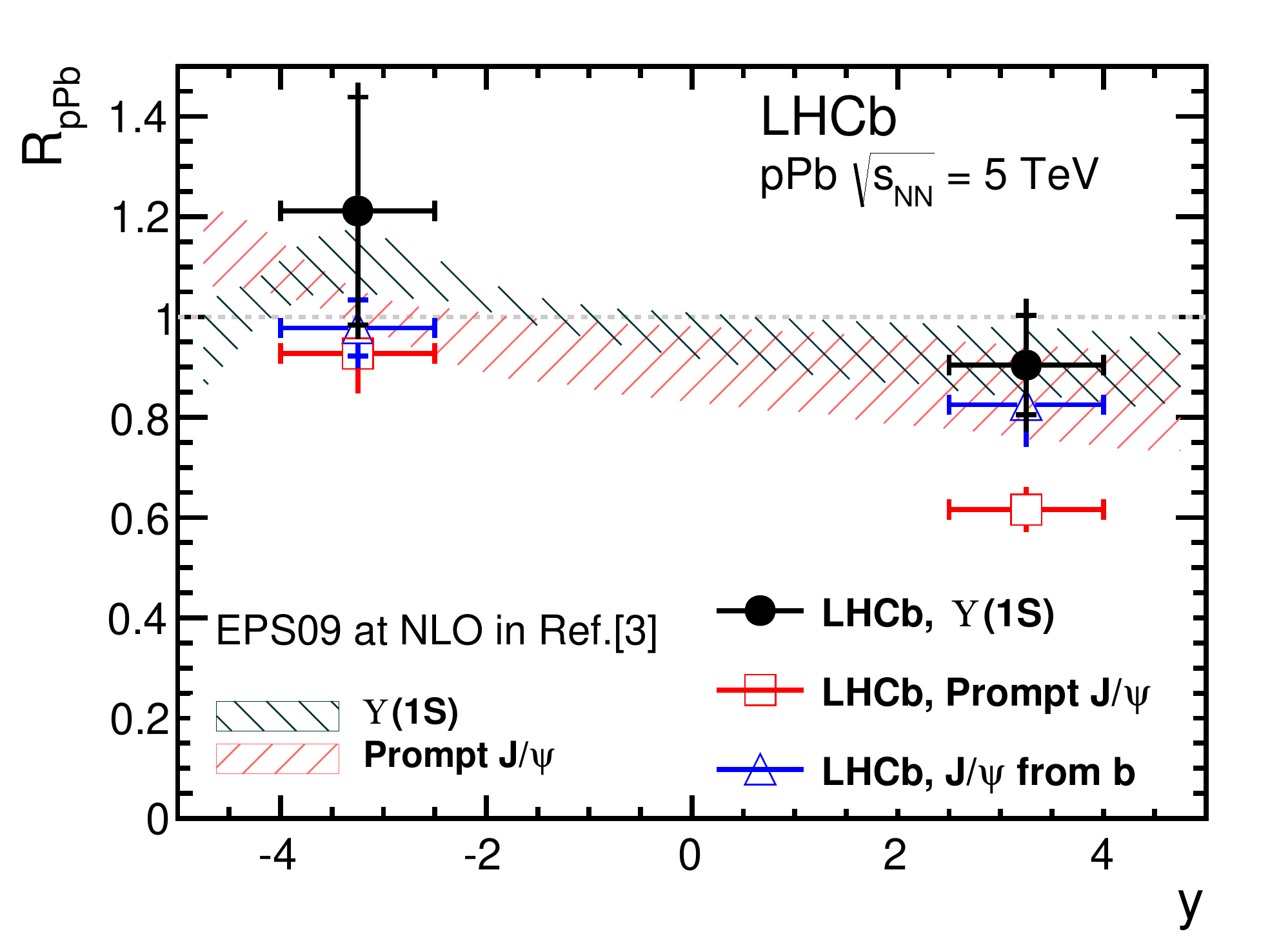}
    \includegraphics[scale=0.35]{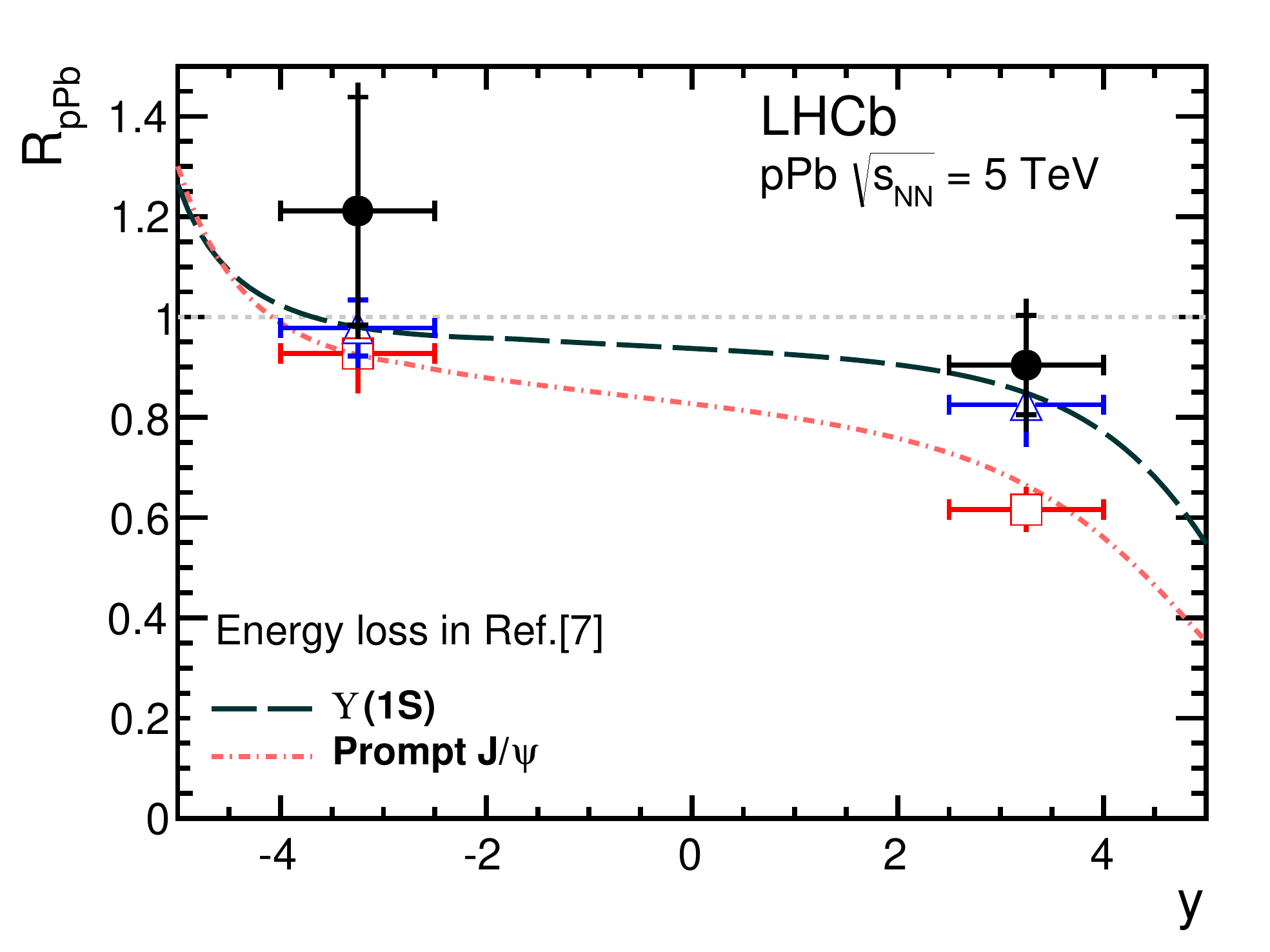}
    \includegraphics[scale=0.35]{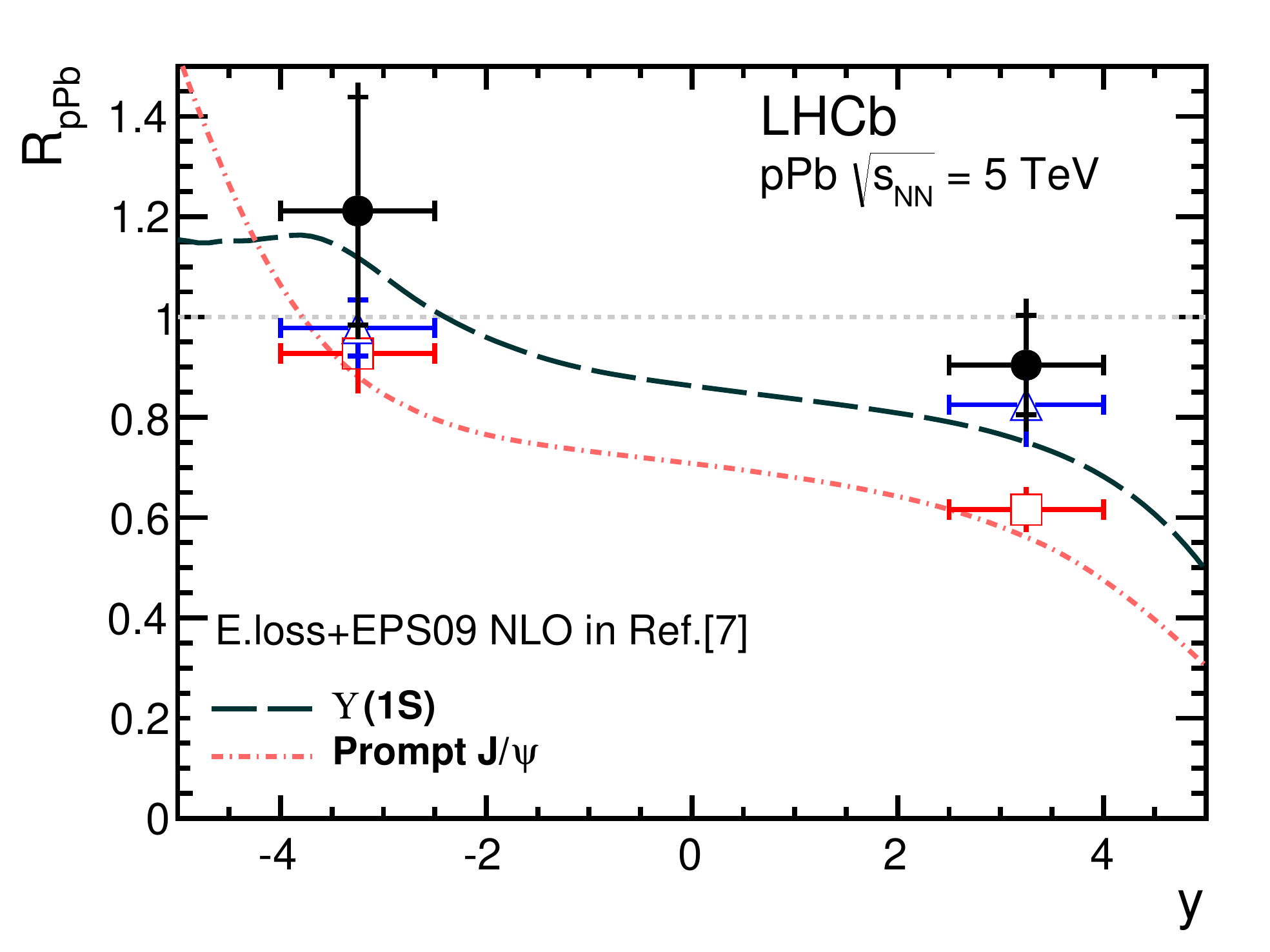}
    \vspace*{-0.5cm}
  \end{center}
  \caption{ \small 
     Nuclear modification factor, $R_{p\mathrm{Pb}}$, 
     compared to other measurements and theoretical predictions.
     The black dots, red squares, and blue triangles indicate 
     the LHCb measurements for $\OneS$ mesons, prompt \jpsi\ mesons, and \jpsi\ from 
     $b$-hadron decays, respectively~\cite{LHCb-PAPER-2013-052}. 
     The inner error bars (delimited by the horizontal lines) 
     show the statistical uncertainties; the outer ones show the statistical 
     and systematic uncertainties added in quadrature.
     The data are compared with theoretical predictions 
     for $\PUpsilon$ and prompt $\jpsi$ mesons
     from different models, one per panel.
     The shaded areas indicate the uncertainties of the theoretical calculations. 
  }
  \label{fig:R_pA}
\end{figure}

%%%%%%%%%%%%%%%%%%%%%%
The nuclear modification factor 
$R_{\pPb}(\sqrt{s_{\mbox{\tiny{\it NN}}}})
\equiv{\sigma_{\pPb}(\sqrt{s_{\mbox{\tiny{\it NN}}}})}/
{(A \times \sigma_{pp}(\sqrt{s_{\mbox{\tiny{\it NN}}}}))}$
is used to study the CNM effects,
where $A$\/ is the atomic mass number of the nucleus
and $\sqrt{s_{\mbox{\tiny{\it NN}}}}$\/ is the centre-of-mass energy of the 
nucleon-nucleon system. The determination of $R_{\pPb}$\/ requires 
the value of the production cross-section in $pp$\/ collisions at $5\tev$,
for which no data is yet available.
Following the same approach as in the  
measurement of $R_{\pPb}$\/ for \jpsi\  mesons \cite{LHCb-PAPER-2013-052},
this cross-section is obtained by a power-law interpolation from previous \lhcb\ 
measurements \cite{LHCb-PAPER-2011-036,LHCb-PAPER-2013-016,LHCb-PAPER-2013-066}
in the range $p_\mathrm{T}<15\gevc$\/ and $2.5<y<4.0$\/ \cite{CommonNote}.
The product of the production cross-section and the dimuon branching fraction
for $\OneS$\/ mesons in $pp$\/ collisions at $5\tev$, with $\pt<15\gevc$\/
and $2.5<y<4.0$, is $\sigma_{pp}\times\BF(1S) =1.12 \pm 0.11{\nb}$,
from which the nuclear modification factors 
$R_{\pPb}$\/ for $\OneS$\/ mesons in the ranges $-4.0<y<-2.5$\/ and 
$2.5<y<4.0$\/ are determined to be
\begin{equation*}
\begin{split}
R_{p\mathrm{Pb}}(\OneS,-4.0<y<-2.5) = 1.21 \pm 0.23 \pm 0.12,\\
R_{p\mathrm{Pb}}(\OneS,\ \xx2.5<y<\ \xx4.0) = 0.90 \pm 0.10 \pm 0.09.\\
\end{split}
\end{equation*} 
Figure~\ref{fig:R_pA} shows the measurement of $R_{\pPb}$\/ for $\OneS$\/ mesons as a function of rapidity.
Relative to the $\OneS$ production in $pp$ collisions, 
the data are consistent with a suppression in the forward region 
and an enhancement due to antishadowing effects in the backward hemisphere.
In the forward region, the data suggest that the suppression of 
$\OneS$\/ production is smaller than that of prompt \jpsi\ production.
The central value of $R_{\pPb}$\/ for $\OneS$\/ mesons is close to 
that for \jpsi from $b$-hadron decays, which reflects the CNM effects on $b$ hadrons.
Within the sizable uncertainties of the current 
measurements, the result agrees with existing theoretical predictions
\cite{Arleo:2012rs,Ferreiro:2011xy,Albacete:2013ei}. The calculations in 
Ref.~\cite{Ferreiro:2011xy} are based on the leading-order CSM,
taking into account the modification of the gluon distribution functions
in the nucleus with the parameterisation EPS09~\cite{Eskola:2009uj}.
The predictions in Ref.~\cite{Albacete:2013ei}
use the next-to-leading-order CEM and the parton shadowing is calculated with 
the EPS09 parameterisation.
Theoretical predictions of the coherent energy loss effect
are provided in Ref. \cite{Arleo:2012rs}, both with and without additional 
parton shadowing effects as parameterised with EPS09.

%%%%% R_{pA}

\begin{figure}[ht]
  \begin{center}
    \includegraphics[scale=0.35]{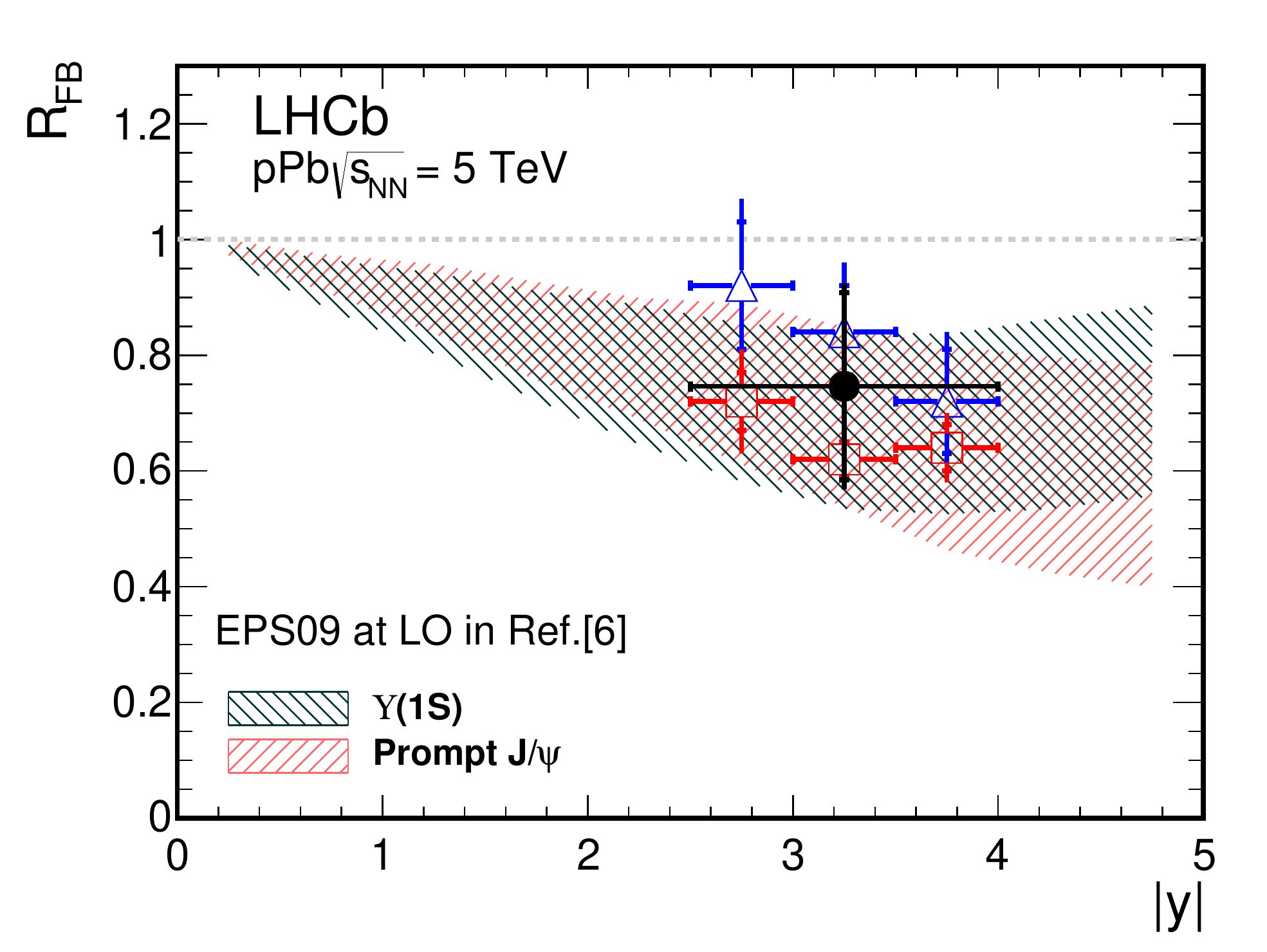}
    \includegraphics[scale=0.35]{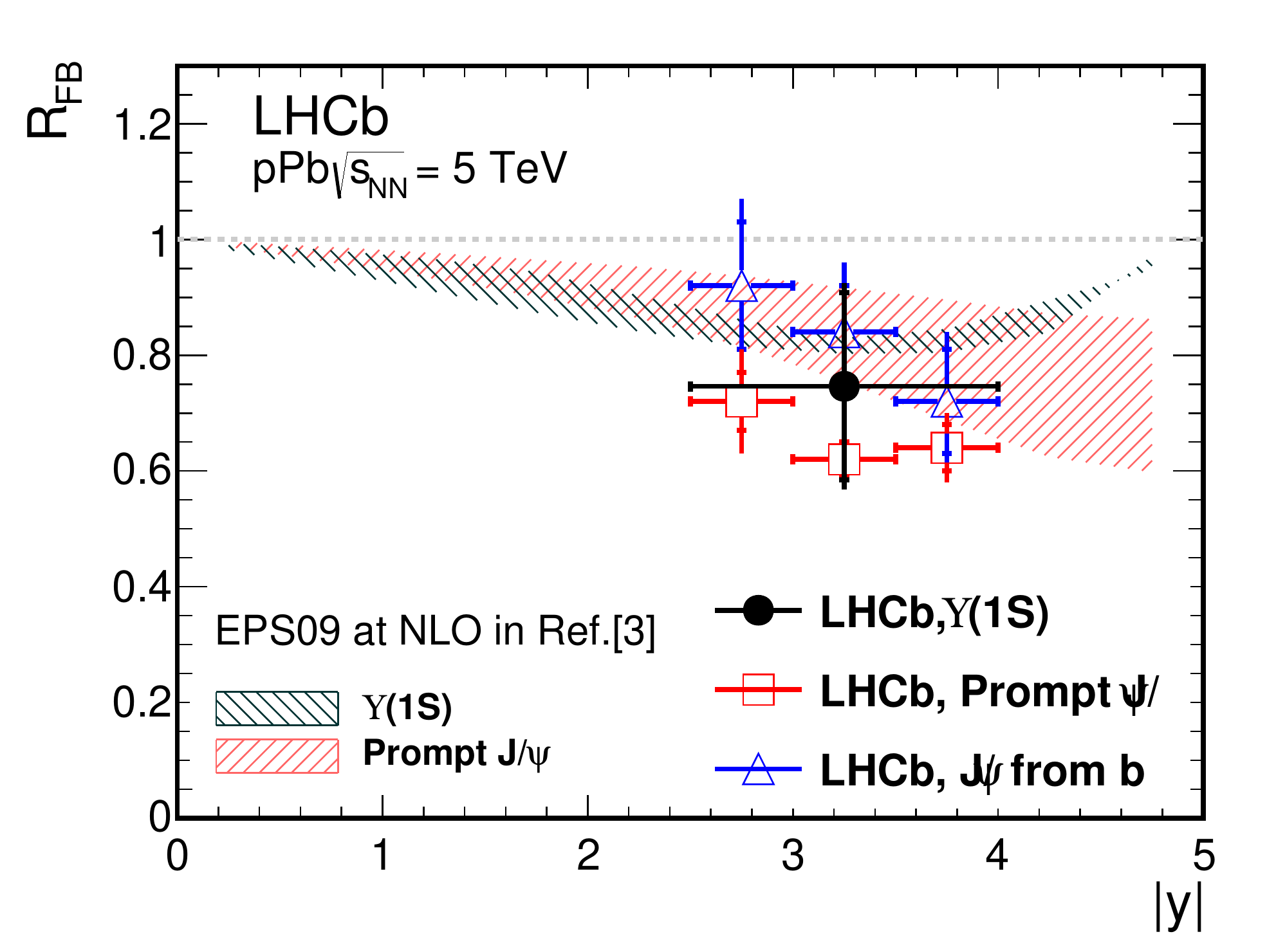}
    \includegraphics[scale=0.35]{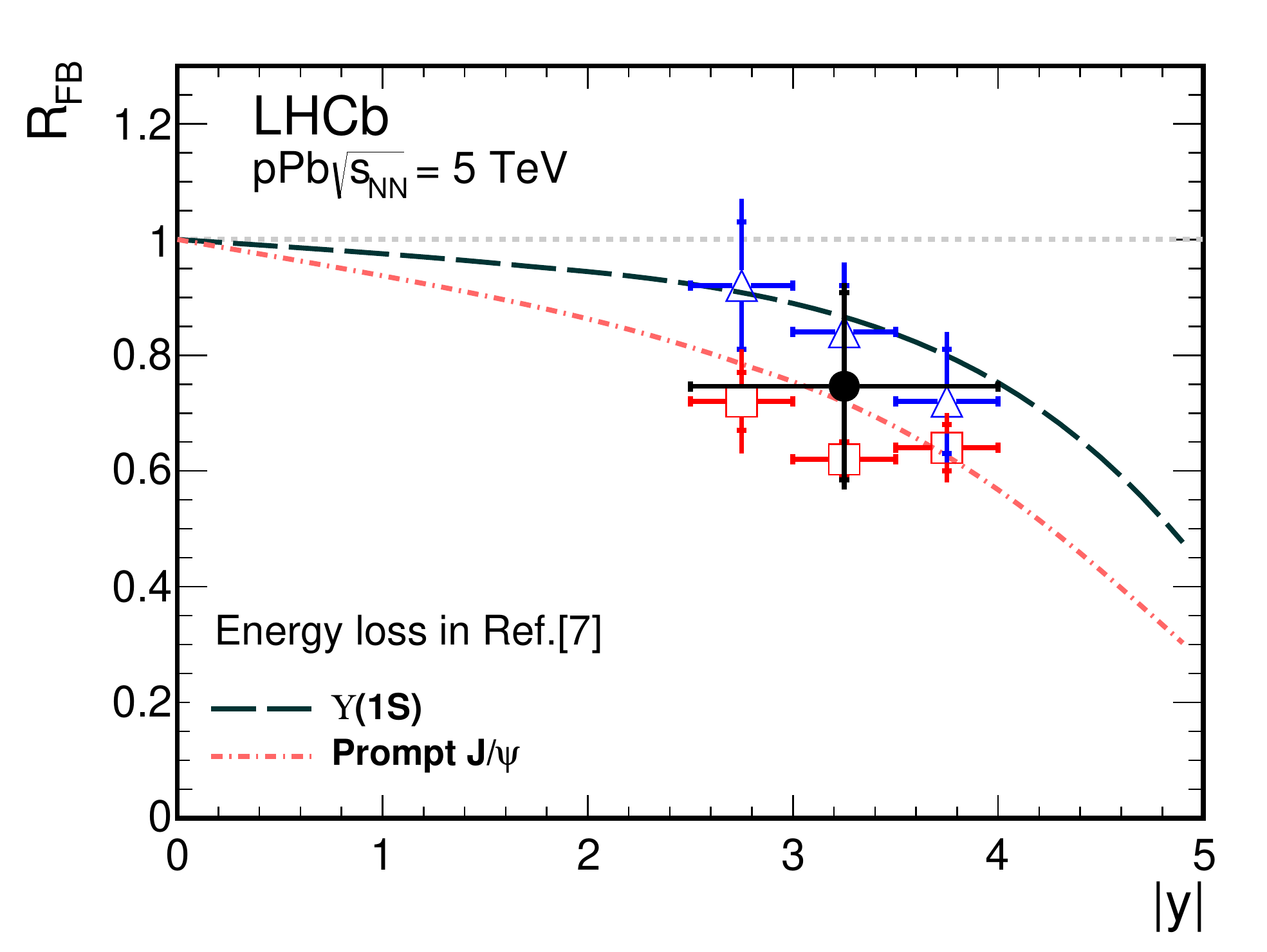}
    \includegraphics[scale=0.35]{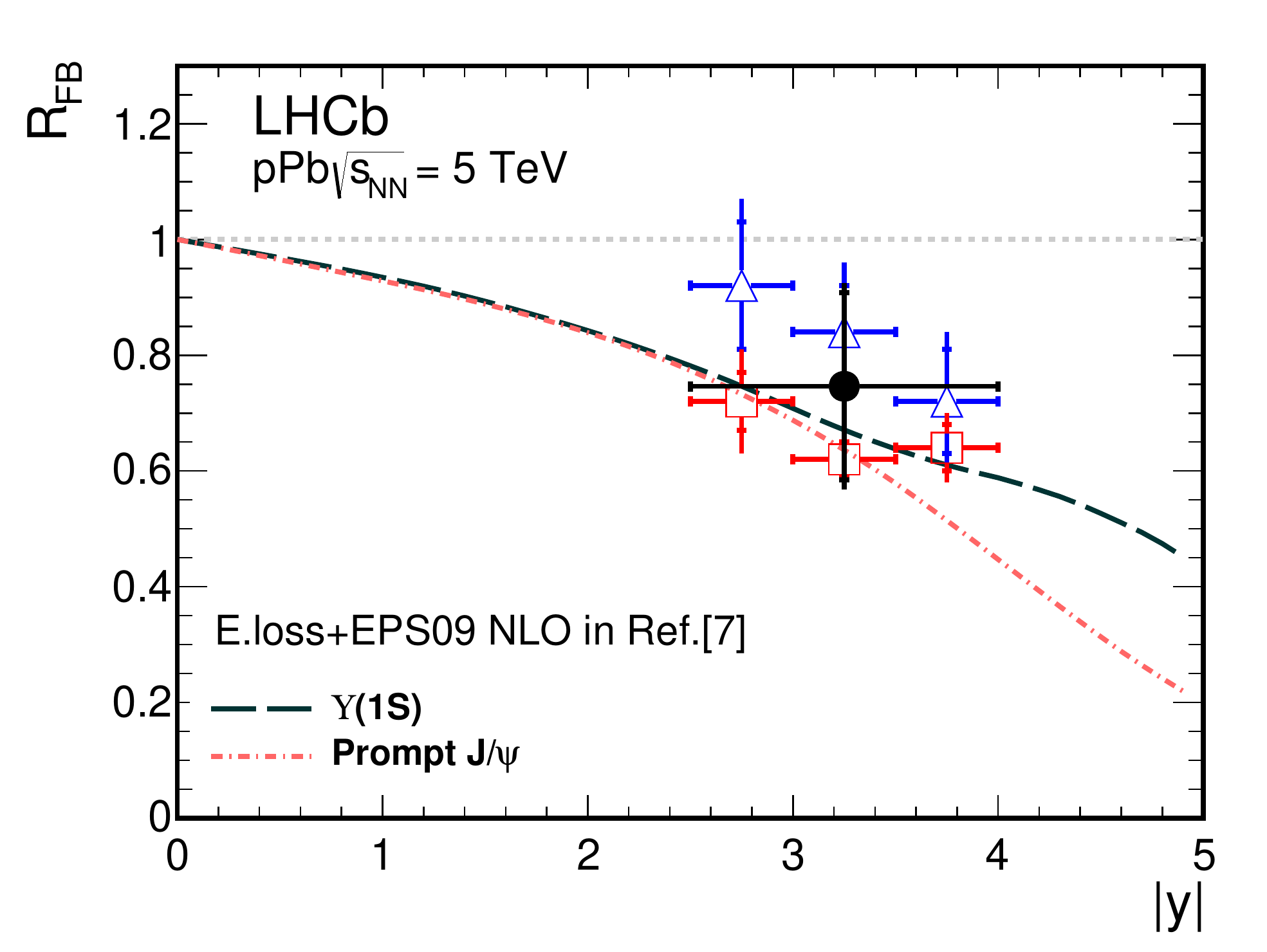}
    \vspace*{-0.8cm}
  \end{center}
  \caption{ \small 
     Forward-backward production ratio, $R_{\mbox{\tiny{FB}}}$, as a function of absolute rapidity. 
     The black dots, red squares, and blue triangles indicate 
     the LHCb measurements for $\OneS$ mesons, prompt \jpsi\ mesons, and \jpsi\ from 
     $b$-hadron decays, respectively~\cite{LHCb-PAPER-2013-052}. 
     The inner error bars (delimited by the horizontal lines) 
     show the statistical uncertainties; the outer ones show the statistical
     and systematic uncertainties added in quadrature.
     The data are compared with theoretical predictions 
     for $\PUpsilon$\/ and prompt $\jpsi$\/ mesons 
     from different models, one per panel.
     The shaded areas indicate the uncertainties of the theoretical calculations. 
  }
\label{fig:R_FB}
\end{figure}

Another observable that characterises CNM effects is the 
forward-backward production ratio, defined as
$R_{\mbox{\tiny{FB}}}(\sqrt{s_{\mbox{\tiny{\it NN}}}},|y|)\equiv
\sigma(\sqrt{s_{\mbox{\tiny{\it NN}}}},+|y|)/\sigma(\sqrt{s_{\mbox{\tiny{\it NN}}}},-|y|)$.
The ratio does not depend on the reference $pp$\/ cross-section,
and part of the experimental and theoretical uncertainties cancel.
The forward-backward production ratio 
of $\OneS$\/ mesons is 
\begin{equation*}
R_\mathrm{\tiny{FB}}(2.5<|y|<4.0)=0.75\pm0.16\pm0.08.
\end{equation*}
Figure~\ref{fig:R_FB} shows the measured value of $R_{\mbox{\tiny{FB}}}$\/ for 
$\OneS$\/ mesons as a function of absolute rapidity, together with the 
theoretical predictions~\cite{Arleo:2012rs,Ferreiro:2011xy,Albacete:2013ei} 
and $R_{\mbox{\tiny{FB}}}$, measured by LHCb,
for prompt $\jpsi$\/ mesons and $\jpsi$\/ from $b$\ hadrons~\cite{LHCb-PAPER-2013-052}. 
Measurements and theoretical predictions agree.

%%%%%% R_{FB}

%%%%%%%%%%%%%%%%
\section{Conclusions}
\label{sec:Conclusions}
The production of $\PUpsilon$\/ mesons is studied in $p\mathrm{Pb}$\/ 
collisions with the {\mbox{LHCb}\xspace}\ detector at a nucleon-nucleon 
centre-of-mass energy $\sqrt{s_{\mbox{\tiny{\it NN}}}}=5\mathrm{\,Te\kern -0.1em V}$\/ 
in the transverse momentum range of $p_\mathrm{T}<15\mathrm{\,Ge\kern -0.1em V}/c$\/
and rapidity range $-5.0<y<-2.5$ and $1.5<y<4.0$.

The nuclear modification factor for the $\OneS$\/ meson is 
determined using the cross-section of $\OneS$\/ production in $pp$\/
collisions at $5\tev$\/ interpolated from previous LHCb measurements.
It is compatible with predictions of a suppression of \OneS\/ production 
with respect to $pp$\/ collisions in the forward region 
and antishadowing effects in the backward region.
The forward-backward production ratio of the $\OneS$\/ is also measured,
and the result is consistent with existing theoretical predictions, 
where the nuclear shadowing effects are taken into account with the EPS09 parameterisation, 
or a coherent energy loss is considered.
%The production ratios of excited $\PUpsilon$ mesons 
%relative to the ground state $\OneS$\/ are measured. 
%Due to the small integrated luminosity of the available data sample, the
%measurements presented here have relatively large uncertainties. 
%More \pPb data are needed for a precise quantitative 
%investigation of cold nuclear matter effects in order to establish a
%reliable baseline for the interpretations of related quark-gluon plasma 
%signatures in nucleus-nucleus collisions and to offer information to constrain
%the parameterisation of theoretical models.
A first measurement of the production ratios of excited $\PUpsilon$ mesons 
relative to the ground state $\PUpsilon$ has been performed. 
Due to the small integrated luminosity of the available data sample, 
the measurements presented here, though very promising, have relatively large uncertainties.
More \pPb data would allow a precise quantitative investigation of cold nuclear matter effects,
to establish a reliable baseline for the interpretations of related quark-gluon plasma signatures
in nucleus-nucleus collisions and constrain the parameterizations of theoretical models.

\vspace{5cm}

% Do not include this in analysis note and conference reports
\section*{Acknowledgements}

\noindent 
We thank F. Arleo, J. P. Lansberg and R. Vogt
for providing us with the theoretical predictions and for the stimulating and helpful discussions.
We express our gratitude to our colleagues in the CERN
accelerator departments for the excellent performance of the LHC. We
thank the technical and administrative staff at the LHCb
institutes. We acknowledge support from CERN and from the national
agencies: CAPES, CNPq, FAPERJ and FINEP (Brazil); NSFC (China);
CNRS/IN2P3 and Region Auvergne (France); BMBF, DFG, HGF and MPG
(Germany); SFI (Ireland); INFN (Italy); FOM and NWO (The Netherlands);
SCSR (Poland); MEN/IFA (Romania); MinES, Rosatom, RFBR and NRC
``Kurchatov Institute'' (Russia); MinECo, XuntaGal and GENCAT (Spain);
SNSF and SER (Switzerland); NASU (Ukraine); STFC and the Royal Society (United
Kingdom); NSF (USA). We also acknowledge the support received from EPLANET, 
Marie Curie Actions and the ERC under FP7. 
The Tier1 computing centres are supported by IN2P3 (France), KIT and BMBF (Germany),
INFN (Italy), NWO and SURF (The Netherlands), PIC (Spain), GridPP (United Kingdom).
We are indebted to the communities behind the multiple open source software packages on which we depend.
We are also thankful for the computing resources and the access to software R\&D tools provided by Yandex LLC (Russia).

\addcontentsline{toc}{section}{References}
\setboolean{inbibliography}{true}
\bibliographystyle{LHCb}
\bibliography{main,LHCb-PAPER,LHCb-CONF,LHCb-DP,pA}

\end{document}